\documentclass[conference,romanappendices,10pt]{IEEEtran}
\IEEEoverridecommandlockouts
\usepackage{algorithm,algorithmic,amsmath,amssymb,amsthm,bbm,cite,color,enumitem,graphicx,microtype,setspace,url}
\usepackage{tikz}
\usepackage[USenglish]{babel}
\usepackage[utf8]{inputenc}
\usepackage[T1]{fontenc}
\usepackage{hyperref}
\usepackage{epstopdf}

{}
{}
{}
{\newtheorem{proposition}{Proposition}}
{}
{}
{}

\newcommand{\f}{\mathbf{f}}
\newcommand{\g}{\mathbf{g}}
\newcommand{\h}{\mathbf{h}}

\newcommand{\p}{\mathbf{p}}

\renewcommand{\v}{\mathbf{v}}
\newcommand{\w}{\mathbf{w}}

\newcommand{\y}{\mathbf{y}}
\newcommand{\z}{\mathbf{z}}

\newcommand{\0}{\mathbf{0}}

\newcommand{\D}{\mathbf{D}}
\newcommand{\E}{\mathbf{E}}

\renewcommand{\H}{\mathbf{H}}
\newcommand{\I}{\mathbf{I}}

\renewcommand{\P}{\mathbf{P}}

\newcommand{\X}{\mathbf{X}}
\newcommand{\Y}{\mathbf{Y}}
\newcommand{\Z}{\mathbf{Z}}

\newcommand{\setB}{\mathcal{B}}
\newcommand{\setC}{\mathcal{C}}

\newcommand{\setG}{\mathcal{G}}

\newcommand{\setK}{\mathcal{K}}

\newcommand{\setN}{\mathcal{N}}

\newcommand{\Real}{\mbox{$\mathbb{R}$}}
\newcommand{\Compl}{\mbox{$\mathbb{C}$}}
\newcommand{\rmF}{\mathrm{F}}

\newcommand{\Diag}{\mathrm{Diag}}

\newcommand{\Exp}{\mathbb{E}}
\newcommand{\herm}{\mathrm{H}}

\renewcommand{\Re}{\mathrm{Re}}

\newcommand{\tran}{\mathrm{T}}

\newcommand{\bs}{\textnormal{\tiny{BS}}}
\newcommand{\dl}{\textnormal{\tiny{DL}}}

\newcommand{\dis}{\textnormal{\tiny{dis}}}
\newcommand{\disgs}{\textnormal{\tiny{dis-gs}}}
\newcommand{\mse}{\mathrm{MSE}}
\newcommand{\sinr}{\mathrm{SINR}}

\newcommand{\ue}{\textnormal{\tiny{UE}}}
\newcommand{\ul}{\textnormal{\tiny{UL}}}
\newcommand{\ulA}{\textnormal{\tiny{UL-1}}}
\newcommand{\ulB}{\textnormal{\tiny{UL-2}}}
\newcommand{\ulC}{\textnormal{\tiny{UL-3}}}

\usepackage{fancyhdr}
\fancypagestyle{firstpage}{
  
  \fancyfoot[C]{{\footnotesize \begin{singlespace} \textcopyright 2022 IEEE. Personal use of this material is permitted. Permission from IEEE must be obtained for all other uses, in any current or future media, including reprinting/republishing this material for advertising or promotional purposes, creating new collective works, for resale or redistribution to servers or lists, or reuse of any copyrighted component of this work in other works.
\end{singlespace}}}}

\title{Distributed Precoding Design for Multi-Group Multicasting in Cell-Free Massive MIMO \vspace{-1mm}}
\author{Bikshapathi Gouda, Italo Atzeni, and Antti Tölli \\
Centre for Wireless Communications, University of Oulu, Finland \\
Emails: \{bikshapathi.gouda, italo.atzeni, antti.tolli\}@oulu.fi \vspace{-1mm}
\thanks{
The work of B.~Gouda and A.~Tölli was supported by the Academy of Finland (318927 6G Flagship). The work of I.~Atzeni was supported by the Marie Skłodowska-Curie Actions (MSCA-IF 897938 DELIGHT) and by the Academy of Finland (336449 Profi6).}\vspace{-1.5mm}}

\begin{document}

\maketitle

\thispagestyle{firstpage}

\begin{abstract}
We consider multi-group multicast precoding designs for cell-free massive multiple-input multiple-output (MIMO) systems. To optimize the transmit and receive beamforming strategies, we focus on minimizing the sum of the maximum mean squared errors (MSEs) over the multicast groups, which is then approximated with the sum MSE to simplify the computation and signaling. We adopt an iterative bi-directional training scheme with uplink and downlink precoded pilots to cooperatively design the multi-group multicast precoders at each base station and the combiners at each user equipment in a distributed fashion. An additional group-specific uplink training resource is introduced, which entirely eliminates the need for backhaul signaling for channel state information (CSI) exchange. We also propose a simpler distributed precoding design based solely on group-specific pilots, which can be useful in the case of scarce training resources. Numerical results show that the proposed distributed methods greatly outperform conventional cell-free massive MIMO precoding designs that rely solely on local CSI.
\end{abstract}

\section{Introduction}

Cell-free massive multiple-input multiple-output (MIMO) is a promising beyond-5G technology that aims at providing a uniform service over its coverage area \cite{Raj20,Int19}. In a cell-free massive MIMO system, a large number of base stations (BSs) are distributed across the network and are connected to a central processing unit (CPU) via backhaul links to exchange the channel state information (CSI) and the user-specific data \cite{Ngo17}. All the BSs jointly serve all the user equipments (UEs) simultaneously, which eliminates the inter-cell interference and enhances the spectral efficiency. Cell-free massive MIMO systems can outperform traditional cellular massive MIMO, and small-cell networks in many practical scenarios \cite{Int19,Ngo17,Bjo20}. Moreover, their performance can be considerably improved by allowing coordination among the BSs \cite{Bjo20}.

The demand for multicasting transmissions is increasing on account of applications such as vehicular communications and augmented/mixed reality \cite{Raj20,Mur19}. In multicasting, the same data is transmitted to a group of UEs with a group-specific precoder. The multicast precoding design framework was presented in \cite{Sid06} and extended to the multi-group setting in \cite{Kar08}. Recently, multi-group multicasting in massive MIMO has been studied in \cite{Sad18,Mah21}. A few works have considered multi-group multicasting via cell-free massive MIMO, which adopts conjugate beamforming for the data transmission \cite{Doa17,Zha19b,Far21,Zho21}. In \cite{Doa17}, equal transmit powers are allocated to all the UEs to avoid any backhaul signaling for CSI exchange. On the other hand, \cite{Zha19b,Far21,Zho21} tackle the max-min fairness problem to compute the transmit powers at the UEs and guarantee the same rate within each multicast group at the cost of extra backhaul signaling. While \cite{Doa17,Zha19b,Far21} consider single-antenna UEs, \cite{Zho21} assumes multi-antenna UEs and antenna-specific pilots for the CSI acquisition.

Considering multi-group multicasting via cell-free massive MIMO with multi-antenna UEs, we aim to cooperatively design the multi-group multicast precoders at each BS and the combiners at each UE in a distributed fashion. To this end, we initially target the minimization of the sum of the maximum mean squared errors (MSEs) over the multicast groups, which we refer to in the following as the \textit{sum-group MSE}. While the sum-group MSE minimization achieves absolute fairness within each multicast group, the resulting distributed precoding design depends on slowly varying dual variables, which results in slow convergence. Instead, we propose to solve a simpler sum MSE minimization problem, which also provides a degree of in-built fairness among all the UEs. This approach provides a good approximation for the sum-group MSE minimization, especially at a medium-to-high signal-to-interference-plus-noise ratio (SINR). Then, we propose a novel distributed precoding design for multi-group multicasting based on iterative bi-directional training \cite{Tol19}. Differently from our previous works \cite{Atz21,Gou20,Atz20} on the unicasting scenario, we introduce an additional group-specific uplink training resource that entirely eliminates the need for backhaul signaling for CSI exchange in multi-group multicasting. We also present a simpler precoding design that relies uniquely on group-specific pilots \cite{Yan13}, which can be useful when the training resources are scarce. Numerical results show that the proposed distributed methods bring substantial gains over conventional cell-free massive MIMO precoding designs that utilize only local CSI.

\section{System Model} \label{sec:SM}

Consider a cell-free massive MIMO system where a set of BSs $\setB \triangleq \{1, \ldots, B\}$ serves a set of UEs $\setK \triangleq \{1,\ldots,K\}$ in the downlink. Each BS and UE is equipped with $M$ and $N$ antennas, respectively. The UEs are divided into a set of non-overlapping multicast groups $\setG \triangleq\{1,\ldots,G\}$. We use $\setK_g$ to denote the set of UEs in group~$g \in \setG$, whereas $g_{k}$ represents the index of the group that contains UE~$k$. The BSs transmit a single data stream to each multicast group, i.e., all the UEs in $\setK_g$ are intended to receive the same data symbol~$d_{g}$. Assuming time division duplex (TDD) mode, let $\H_{b,k} \in \Compl^{M \times N}$ denote the uplink channel matrix between UE~$k \in \setK$ and BS~$b \in \setB$, and let $\w_{b,g} \in \Compl^{M \times 1}$ be the BS-specific precoder used by BS~$b$ for group~$g$. We use $\H_{k} \triangleq [\H_{1,k}^{\tran}, \ldots, \H_{B,k}^{\tran}]^{\tran} \in \Compl^{B M \times N}$ and $\w_{g} \triangleq [\w_{1,g}^{\tran}, \ldots, \w_{B,g}^{\tran}]^{\tran} \in \Compl^{B M \times 1}$ to denote the aggregated uplink channel matrix of UE~$k$ and the aggregated precoder used for group~$g$, respectively, which imply $\H_{k}^{\herm} \w_{g} = \sum_{b \in \setB} \H_{b,k}^{\herm} \w_{b,g}$. We assume the per-BS transmit power constraints $\sum_{g \in \setG} \|\w_{b,g}\|^{2} \leq \rho_{\bs},~\forall b \in \setB$, where $\rho_{\bs}$ denotes the maximum transmit power at each BS. Hence, the received signal at UE~$k$ is given by
\begin{align}
\nonumber \y_{k} \triangleq \sum_{b \in \setB} \! \H_{b,k}^{\herm} \w_{b,g_{k}} d_{g_{k}} \! + \! \sum_{\bar{g} \neq g_{k}} \sum_{b \in \setB} \! \H_{b,k}^{\herm} \w_{b,\bar{g}} d_{\bar{g}} \! + \! \z_{k} \! \in \! \Compl^{N \times 1},
\end{align}
\vspace{-6mm}
\begin{align}
\label{eq:y_k}
\end{align}
where $\z_{k} \in \setC \setN (\0, \sigma_{\ue}^{2} \I_{N})$ is the additive white Gaussian noise (AWGN). Upon receiving $\y_{k}$, UE~$k$ applies the combiner $\v_{k} \in \Compl^{N \times 1}$ to obtain a soft estimate of $d_{g}$ and the resulting SINR can be expressed as
\begin{align} \label{eq:SINR_k}
\sinr_{k} & \triangleq \frac{|\sum_{b \in \setB} \v_{k}^{\herm} \H_{b,k}^{\herm} \w_{b,g_{k}}|^{2}}{\sum_{\bar g \ne g_{k}} |\sum_{b \in \setB} \v_{k}^{\herm} \H_{b,k}^{\herm} \w_{b,\bar{g}}|^{2} + \sigma_{\ue}^{2} \| \v_{k} \|^{2}}.
\end{align}
Finally, the sum of the rates over the multicast groups, which we refer to in the following as the \textit{sum-group rate}, is given by
\begin{align} \label{eq:R}
R \triangleq \sum_{g \in \setG} \min_{k \in \setK_g} \log_{2}(1 + \sinr_{k}) \quad \textrm{[bps/Hz]}.
\end{align}
Note that \eqref{eq:R} represents an upper bound on the system performance that is attained in presence of perfect global CSI.

In Section~\ref{sec:distr}, we propose distributed precoding designs based on iterative bi-directional training to cooperatively~compute the multi-group multicast precoders at each BS and the combiners at each UE. We also present the centralized precoding design along with the problem formulation~in~Section~\ref{sec:problem}.

\subsection{Bi-Directional Training and Channel Estimation} \label{sec:SM_est}

\vspace{-0.5mm}

The proposed distributed precoding designs rely on iterative bi-directional training, where uplink and downlink precoded pilots are transmitted at each iteration \cite{Tol19}. On the other hand, the centralized precoding design requires each BS to estimate the antenna-specific uplink channels. We now describe the different types of pilot-aided channel estimation that will be used in Sections~\ref{sec:problem} and~\ref{sec:distr}.

\smallskip

\textit{\textbf{Antenna-specific uplink channel estimation.}} The estimation of the channel matrix $\H_{b,k}$ involves $N$ antenna-specific uplink pilots for UE~$k$. Let $\P_{k} \in \Compl^{\tau \times N}$ be the pilot matrix assigned to UE~$k$, with $\|\P_{k}\|_{\rmF}^{2} = \tau N$ and $\P_{k}^{\herm} \P_{k} = \tau \I_{N}$. Moreover, let $\rho_{\ue}$ denote the maximum transmit power at each UE. Each UE~$k$ synchronously transmits its pilot matrix, i.e.,
\begin{align} \label{eq:X_k_ul}
\X_{k}^{\ul} \triangleq \sqrt{\beta^{\ul}} \P_{k}^{\herm} \in \Compl^{N \times \tau},
\end{align}
where the power scaling factor $\beta^{\ul} \triangleq \frac{\rho_{\ue}}{N}$ ensures that $\X_{k}^{\ul}$ complies with the UE transmit power constraint. Then, the received signal at BS~$b$ is given by
\begin{align}
\Y_{b}^{\ul} & \ \triangleq \sum_{k \in \setK} \H_{b,k} \X_{k}^{\ul} + \Z_{b}^{\ul} \\
\label{eq:Y_b_ul} & = \ \sqrt{\beta^{\ul}} \sum_{k \in \setK} \H_{b,k} \P_{k}^{\herm} + \Z_{b}^{\ul} \in \Compl^{M \times \tau},
\end{align}
where \hfill $\Z_{b}^{\ul}$ \hfill is \hfill the \hfill AWGN \hfill with \hfill i.i.d. \hfill $\setC \setN (0, \sigma_{\bs}^{2})$ \hfill elements.

\noindent Finally, the least-squares (LS) estimate of $\H_{b,k}$ is obtained as
\begin{align}
\hat{\H}_{b,k} & \triangleq \frac{1}{\tau \sqrt{\beta^{\ul}}} \Y_{b}^{\ul} \P_{k} \\
\label{eq:H_bk_hat_orth} & = \H_{b,k} + \frac{1}{\tau} \sum_{\bar{k}  \ne k} \H_{b,\bar{k}} \P_{\bar{k}}^{\herm} \P_{k} + \frac{1}{\tau \sqrt{\beta^{\ul}}} \Z_{b}^{\ul} \P_{k}.
\end{align}

\smallskip

\textit{\textbf{UE-specific effective uplink channel estimation.}} Let $\h_{b,k} \triangleq \H_{b,k} \v_{k} \in \Compl^{M \times 1}$ denote the effective uplink channel between UE~$k$ and BS~$b$, and let $\p_{k} \in \Compl^{\tau \times 1}$ be the pilot assigned to UE~$k$, with $\|\p_{k}\|^{2} = \tau$. Each UE~$k$ synchronously transmits its pilot $\p_{k}$ using its combiner $\v_{k}$ as precoder, i.e.,
\begin{align} \label{eq:X_k_ul1}
\X_{k}^{\ulA} \triangleq \sqrt{\beta^{\ulA}} \v_{k} \p_{k}^{\herm} \in \Compl^{N \times \tau},
\end{align}
where the power scaling factor $\beta^{\ulA}$ ensures that $\X_{k}^{\ulA}$ complies with the UE transmit power constraint. Then, the received signal at BS~$b$ is given by
\begin{align}
\Y_{b}^{\ulA} & \triangleq \sum_{k \in \setK} \H_{b,k} \X_{k}^{\ulA} + \Z_{b}^{\ulA} \\
\label{eq:Y_b_ul1} & = \sqrt{\beta^{\ulA}} \sum_{k \in \setK} \h_{b,k} \p_{k}^{\herm} + \Z_{b}^{\ulA} \in \Compl^{M \times \tau},
\end{align}
where $\Z_{b}^{\ulA}$ is the AWGN with i.i.d. $\setC \setN (0, \sigma_{\bs}^{2})$ elements. Finally, the LS estimate of $\h_{b,k}$ is obtained as 
\begin{align}
\hat{\h}_{b,k} & \triangleq \frac{1}{\tau \sqrt{\beta^{\ulA}}} \Y_{b}^{\ulA} \p_{k} \\
\label{eq:h_bk_hat} & = \h_{b,k} + \frac{1}{\tau} \sum_{\bar{k} \ne k} \h_{b,\bar{k}} \p_{\bar{k}}^{\herm} \p_{k} + \frac{1}{\tau \sqrt{\beta^{\ulA}}} \Z_{b}^{\ulA} \p_{k}.
\end{align}

\smallskip

\textit{\textbf{Group-specific effective uplink channel estimation.}} Let $\f_{b,g} \triangleq \sum_{ k \in \setK_g} \H_{b,k} \v_{k} \in \Compl^{M \times 1}$ denote the effective uplink channel between $\setK_{g}$ and BS~$b$, and let $\p_{g} \in \Compl^{\tau \times 1}$ be the pilot assigned to group~$g$, with $\|\p_{g}\|^{2} = \tau$. Each UE~$k$ synchronously transmits its pilot $\p_{g_{k}}$ using its combiner $\v_{k}$ as precoder, i.e.,
\begin{align} \label{eq:X_g_ul2}
\X_{k}^{\ulB} \triangleq \sqrt{\beta^{\ulB}} \v_{k} \p_{g_{k}}^{\herm} \in \Compl^{N \times \tau},
\end{align}
where the power scaling factor $\beta^{\ulB}$ ensures that $\X_{k}^{\ulB}$ complies with the UE transmit power constraint. Then, the received signal at BS~$b$ is given by
\begin{align}
\Y_{b}^{\ulB} & \triangleq \sum_{k \in \setK} \H_{b,k} \X_{k}^{\ulB} + \Z_{b}^{\ulB} \\
\label{eq:Y_b_ul2} & = \sqrt{\beta^{\ulB}} \sum_{g \in \setG} \f_{b,g} \p_{g}^{\herm} + \Z_{b}^{\ulB} \in \Compl^{M \times \tau},
\end{align}
where $\Z_{b}^{\ulB}$ is the AWGN with i.i.d. $\setC \setN (0, \sigma_{\bs}^{2})$ elements. Finally, the LS~estimate~of~$\f_{b,g}$~is~obtained~as
\begin{align}
\hat{\f}_{b,g} & \triangleq \frac{1}{\tau \sqrt{\beta^{\ulB}}} \Y_{b}^{\ulB} \p_{g} \\
\label{eq:f_bk_hat} & = \f_{b,g} + \frac{1}{\tau} \sum_{\bar{g} \ne g} \f_{b,\bar{g}} \p_{\bar{g}}^{\herm} \p_{g} + \frac{1}{\tau \sqrt{\beta^{\ulB}}} \Z_{b}^{\ulB} \p_{g}.
\end{align}

\smallskip

\textit{\textbf{Effective downlink channel estimation.}} Let $\g_{k} \triangleq \sum_{b \in \setB} \H_{b,k}^{\herm} \w_{b,g} \in \Compl^{N \times 1}$ denote the effective downlink channel between all the BSs and UE~$k$. Each BS~$b$ synchronously transmits a superposition of the pilots $\{\p_{g}\}_{g \in \setG}$ after precoding them with the corresponding group-specific precoders $\{\w_{b,g}\}_{g \in \setG}$, i.e.,
\begin{align} \label{eq:X_b_dl}
\X_{b}^{\dl} \triangleq \sum_{g \in \setG} \w_{b,g} \p_{g}^{\herm} \in \Compl^{M \times \tau}.
\end{align}
Then, the received signal at UE~$k$ is given by 
\begin{align}
\Y_{k}^{\dl} & \triangleq \sum_{b \in \setB} \H_{b,k}^{\herm} \X_{b}^{\dl} + \Z_{k}^{\dl} \\
\label{eq:Y_k_dl} & = \sum_{b \in \setB} \sum_{g \in \setG} \H_{b,k}^{\herm} \w_{b,g} \p_{g}^{\herm} + \Z_{k}^{\dl} \in \Compl^{N \times \tau},
\end{align}
where $\Z_{k}^{\dl}$ is the AWGN with i.i.d. $\setC \setN (0, \sigma_{\ue}^{2})$ elements. Finally, the LS estimate of $\g_{k}$ is obtained as 
\begin{align}
\hat{\g}_{k} & \triangleq \frac{1}{\tau} \Y_{k}^{\dl} \p_{g_{k}} \\
\label{eq:g_k_hat} & = \g_{k} + \frac{1}{\tau} \sum_{b \in \setB} \sum_{\bar{g} \ne g_{k}} \H_{b,k}^{\herm} \w_{b,\bar{g}} \p_{\bar{g}}^{\herm} \p_{g_{k}} + \frac{1}{\tau} \Z_{k}^{\dl} \p_{g_{k}}.
\end{align}

\section{Problem Formulation} \label{sec:problem}

In this section, we present the problem formulation for the proposed multi-group multicast precoding design focusing on the centralized method, where the aggregated precoders are computed at the CPU. In doing so, we first target the sum-group MSE minimization in Section~\ref{sec:problem_sum-groupMSE} and then propose to solve a simpler sum MSE minimization problem in Section~\ref{sec:problem_sum-groupMSE}.

\subsection{Sum-Group MSE Minimization} \label{sec:problem_sum-groupMSE}

The sum-group MSE minimization achieves absolute fairness within each multicast group through the min-max MSE criterion and can be expressed as
\begin{align} \label{eq:probForHi}
\begin{array}{cl}
\displaystyle \underset{{\{\w_g, \v_{k}\}}}{\mathrm{minimize}} & \displaystyle \sum_{g \in \setG} \max_{k \in \setK_g} \mse_k \\
\mathrm{s.t.} & \displaystyle \sum_{g \in \setG} \| \E_{b} \w_{g}\|^{2} \leq \rho_{\bs}, \quad \forall b \in \setB,
\end{array}
\end{align}
where $\mse_k$ is defined as
\begin{align}
\mse_{k} & \triangleq \Exp\big[ |\v_{k}^{\herm} \y_{k} - d_{g_k}|^{2} \big] \\
&= \label{eq:MSE_k}
\sum_{g \in \setG} | \v_{k}^{\herm} \H_{k}^{\herm} \w_{g} |^{2} - 2 \Re [ \v_{k}^{\herm} \H_{k}^{\herm} \w_{g_k}] \nonumber \\
& \phantom{=} \ + \sigma_{\ue}^{2} \| \v_{k} \|^{2} + 1,
\end{align}
and $\E_{b} \in \Real^{M \times B M}$ is such that $\E_b \w_g = \w_{b,g}$. The problem in \eqref{eq:probForHi} is convex with respect to either the precoders or the combiners. Hence, we use \textit{alternating optimization}, whereby the precoders are optimized for fixed combiners and vice versa in an iterative best-response fashion. Before describing each step of the alternating optimization, let us define $t_g \triangleq \max_{k \in \setK_g} \mse_k$ and rewrite \eqref{eq:probForHi} as
\begin{align} \label{eq:probFor}
\begin{array}{cl}
\displaystyle \underset{{\{t_g, \w_g, \v_{k}\}}}{\mathrm{minimize}} & \displaystyle \sum_{g \in \setG} t_g \\
\mathrm{s.t.} & \displaystyle \mse_{k} \le t_{g}, \quad \forall k \in \setK_g,~\forall g \in \setG \\
& \displaystyle \sum_{g \in \setG} \| \E_{b} \w_{g}\|^{2} \leq \rho_{\bs}, \quad \forall b \in \setB.
\end{array}
\end{align}

\smallskip

\textit{\textbf{Optimization of the combiners.}} For a fixed set of precoders, the combiners $\{\v_{k}\}_{k \in \setK}$ are obtained by solving
\begin{align} \label{eq:probUE}
\begin{array}{cl}
\displaystyle \underset{{\{t_g, \v_{k}\}}}{\mathrm{minimize}} & \displaystyle \sum_{g \in \setG} t_g  \\
\mathrm{s.t.} &  \displaystyle \mse_{k} \le t_{g},  \quad  \forall k \in \setK_g,~\forall g \in \setG.
\end{array}
\end{align}

\noindent Specifically, the optimal $\v_{k}$ can be obtained by computing the stationary point of the Lagrangian of \eqref{eq:probUE}, which yields 
\begin{align}\label{eq:uebf}
\v_{k} = \bigg( \sum_{g \in \setG}\H_{k}^{\herm} \w_{g} \w_{g}^{\herm} \H_{k} + \sigma_{\ue}^{2} \I_{N} \bigg)^{-1}\H_{k}^{\herm} \w_{g_k}.
\end{align}

\smallskip

\textit{\textbf{Optimization of the precoders.}} For a fixed set of combiners, the precoders $\{\w_{g}\}_{g \in \setG}$ are obtained by solving
\begin{align} \label{eq:probBS}
\begin{array}{cl}
\displaystyle \underset{{\{t_g, \w_g \}}}{\mathrm{minimize}} &  \displaystyle \sum_{g \in \setG} t_g \\
\mathrm{s.t.}  & \displaystyle \mse_{k} \le t_{g},  \quad  \forall k \in \setK_g,~\forall g \in \setG \\
& \displaystyle \sum_{g \in \setG} \| \E_{b} \w_{g}\|^{2} \leq \rho_{\bs}, \quad \forall b \in \setB. 
\end{array}
\end{align}
Specifically, the optimal $\w_{g}$ can be obtained by computing the stationary point of the Lagrangian of \eqref{eq:probBS}, which yields
\begin{align}\label{eq:bsbf}
\w_{g} = \bigg( \! \sum_{k \in \setK} \! \nu_{k} \H_{k} \v_{k} \v_{k}^{\herm} \H_{k}^{\herm} \! + \! \sum_{b \in \setB} \!\lambda_{b} \E_{b}^{\herm} \E_{b} \! \bigg)^{-1} \! \sum_{k \in \setK_g} \! \nu_{k} \H_{k} \v_{k},
\end{align}
where $\nu_k$ and $\lambda_b$ are the dual variables corresponding to the first and second constraints in \eqref{eq:probBS}, respectively. Note that $\nu_k$ and $\lambda_b$ can be updated using the sub-gradient and ellipsoid methods, respectively.

\smallskip

The combiner in \eqref{eq:uebf} can be computed locally at each UE. However, the local computation of the precoder in \eqref{eq:bsbf} at each BS requires BS-specific CSI from all the other BSs \cite{Atz21}. Moreover, the sub-gradient update of the dual variables $\{\nu_{k}\}_{k \in \setK}$ significantly slows down the convergence. To simplify the distributed precoding design, we thus propose to replace the sum-group MSE minimization with the sum MSE minimization, as described next.

\subsection{Sum MSE Minimization} \label{sec:problem_sumMSE}

To circumvent the shortcomings of the sum-group MSE minimization, we propose to tackle the (weighted) sum MSE minimization, which can be expressed as
\begin{align} \label{eq:EqvprobForHi}
\begin{array}{cl}
\displaystyle \underset{{\{\w_g, \v_{k}\}}}{\mathrm{minimize}} & \displaystyle \sum_{k \in \setK} \mu_k\mse_k \\
\mathrm{s.t.} & \displaystyle \sum_{g \in \setG} \| \E_{b} \w_{g}\|^{2} \leq \rho_{\bs},  \quad \forall b \in \setB, 
\end{array}
\end{align}
where $\mu_k$ is the weight of UE~$k$. This choice stems from the fact that \eqref{eq:EqvprobForHi} provides a degree of in-built fairness among all the UEs, especially at high SINR. More specifically, at high SINR, the sum MSE minimization corresponds to maximizing the minimum SINR across all the UEs and well approximates the sum-group MSE minimization in \eqref{eq:probForHi}. This is formalized in Proposition~\ref{pre:1}, whose proof is omitted due to the space limitations and will be presented in the longer version of this paper.

\begin{proposition}\label{pre:1}
At high SINR, both the sum-group MSE minimization in \eqref{eq:probForHi} and the sum MSE minimization in \eqref{eq:EqvprobForHi} maximize the minimum SINR across all the UEs.
\end{proposition}

Note that, regardless of the high-SINR assumption, \eqref{eq:EqvprobForHi} becomes equivalent to \eqref{eq:probForHi} if $\mu_k = \nu_k,~\forall k$ at each iteration of the alternating optimization. However, optimally tuning $\{\mu_k\}_{k \in \setK}$ at each iteration leads to the same complexity and signaling requirements as the sum-group MSE minimization. To simplify the distributed precoding design, we fix $\{\mu_k\}_{k \in \setK}$ to the same value at all the iterations. Though slightly suboptimal, this approach leads to much simpler computation and signaling as well as faster convergence. In this context, the optimal $\w_{g}$ can be obtained by computing the stationary point of the Lagrangian of \eqref{eq:EqvprobForHi} for a fixed set of combiners, which yields
\begin{align}\label{eq:bsbf_mse}
\w_{g} = \bigg( \! \sum_{k \in \setK} \! \mu_{k} \H_{k} \v_{k} \v_{k}^{\herm} \H_{k} ^{\herm} \! + \! \sum_{b \in \setB} \! \lambda_{b} \E_{b}^{\herm} \E_{b} \! \bigg)^{-1} \! \sum_{k \in \setK_g} \! \mu_{k} \H_{k} \v_{k}.
\end{align}
Furthermore, the optimal $\v_k$ of \eqref{eq:EqvprobForHi} for a fixed set of precoders corresponds to \eqref{eq:uebf}. Hereafter, we consider the sum MSE minimization to design the multi-group multicast precoders.

At this stage, we briefly illustrate the centralized precoding design with pilot-aided channel estimation. The algorithm begins with the antenna-specific uplink channel estimation described in Section~\ref{sec:SM_est}, whereby each BS~$b$ obtains $\{\hat \H_{b,k}\}_{k \in \setK}$ and forwards them to the CPU via backhaul signaling. Then, the CPU computes the combiners $\{\v_k\}_{k \in \setK}$ and the aggregated precoders $\{\w_g\}_{g \in \setG}$ via alternating optimization by replacing $\H_k$ with $\hat \H_k \triangleq [\hat \H_{1,k}^{\tran}, \ldots, \hat \H_{B,k}^{\tran}]^{\tran}$ in \eqref{eq:uebf} and \eqref{eq:bsbf_mse}, respectively. After convergence, the resulting BS-specific precoders are fed back to the corresponding BSs via backhaul signaling. Finally, the effective downlink channel estimation described in Section~\ref{sec:SM_est} is performed and each UE~$k$ computes its final combiner as
\begin{align}\label{eq:rxmmse}
\v_{k} & = \big(\Y_k^{\dl} (\Y_k^{\dl})^{\herm}\big)^{-1} \Y_k^{\dl} \p_{g_{k}},
\end{align}
which is equal to \eqref{eq:uebf} for perfect channel estimation.

\section{Distributed Precoding Design} \label{sec:distr}

In the distributed precoding design, the optimal $\w_{b,g}$ can be obtained by computing the stationary point of the Lagrangian of \eqref{eq:EqvprobForHi} with respect to $\w_{b,g}$, which yields
\begin{align}\label{eq:dis_bsbf}
\w_{b,g} & = \bigg( \! \sum_{k \in \setK} \! \mu_{k} \H_{b,k} \v_{k} \v_{k}^{\herm} \H_{b,k} ^{\herm} \! + \! \lambda_{b} \I_M \! \bigg)^{-1} \! \bigg( \! \sum_{k \in \setK_g} \! \mu_{k} \H_{b,k} \v_{k} \nonumber \\
& \phantom{=} \ - \underbrace{\sum_{\bar b \ne b} \sum_{k \in \setK} \mu_{k}\H_{b,k}\v_{k} \v_{k}^{\herm} \H_{\bar b,k} ^{\herm}\w_{\bar b,g}}_{\textnormal{cross terms}}\bigg).
\end{align}
This can be computed locally at each BS upon receiving the cross terms from all the other BSs via backhaul signaling. Furthermore, ensuring global convergence of the distributed precoding design requires an iterative best-response update as
\begin{align} \label{eq:w_bk_i}
\w_{b,g}^{(i)} & = \w_{b,g}^{(i-1)} + \alpha \underbrace{(\w_{b,g}-\w_{b,g}^{(i-1)})}_{\triangleq \w_{b,g}^{\star}},
\end{align}
where $i$ is the iteration index and $\alpha \in (0,1]$ determines the trade-off between convergence speed and accuracy \cite{Atz21}. Assuming \hfill a \hfill single-iteration \hfill backhaul \hfill delay \hfill to \hfill exchange \hfill the

\noindent cross terms among the BSs, $\w_{b,g}^{\star}$ in \eqref{eq:w_bk_i} can be written as
\begin{align} \label{eq:w_bg_*}
\hspace{-3mm} \w_{b,g}^{\star} & = \bigg( \! \sum_{k \in \setK} \! \mu_{k} \H_{b,k} \v_{\bar k} \v_{k}^{\herm} \H_{b,k}^{\herm} \! + \! \lambda_{b} \I_M \! \bigg)^{-1} \! \bigg( \! \sum_{k \in \setK_g} \! \mu_{k} \H_{b,k} \v_{k} \nonumber \\
& \phantom{=} \ - \sum_{\bar b \in \setB} \sum_{k \in \setK} \! \mu_{k} \H_{b,k} \v_{k} \v_{k}^{\herm} \H_{\bar b,k}^{\herm}\w_{\bar b,g}^{(i-1)} \! - \! \lambda_{b} \w_{b,g}^{(i-1)} \! \bigg). \hspace{-2mm}
\end{align}

\begin{figure*}[ht]
\addtocounter{equation}{+3}
\begin{align}
\label{eq:bsbflocal} \w_{b,g}^{\dis} & = \bigg(\Y_{b}^{\ulA} \D_{\mu} ({\Y_{b}^{\ulA}})^{\herm} + \tau({\beta^{\ulA}}\lambda_{b} - \sigma_{\bs}^2)\I_M\bigg)^{-1} \bigg( \sqrt{\beta^{\ulA}} \sum_{k \in \setK_g} \mu_k\Y_{b}^{\ulA}\p_k - \frac{{\beta^{\ulA}}}{\sqrt{\beta^{\ulC}}}\Y_{b}^{\ulC}\p_g^{\herm} - {{\beta^{\ulA}}} \tau\lambda_b \w_{b,g}^{(i-1)}\bigg)
\end{align}
\addtocounter{equation}{0}
\hrulefill \vspace{-3mm}
\end{figure*}
\begin{figure*}[ht]
\addtocounter{equation}{0}
\begin{align}
\label{eq:bsbflocalgs} \w_{b,g}^{\disgs} &= \bigg(\Y_{b}^{\ulB}({\Y_{b}^{\ulB}})^{\herm} + \tau({\beta^{\ulB}}\lambda_{b} - \sigma_{\bs}^2)\I_M\bigg)^{-1} \bigg( \sqrt{\beta^{\ulB}} \Y_{b}^{\ulB}\p_g - \frac{{\beta^{\ulB}}}{\sqrt{\beta^{\ulC}}}\Y_{b}^{\ulC}\p_g^{\herm} - {{\beta^{\ulB}}} \tau\lambda_b \w_{b,g}^{(i-1)}\bigg) 
\end{align}
\addtocounter{equation}{-5}
\hrulefill \vspace{-5mm}
\end{figure*}
\subsection{Best-Response Distributed Precoding Design} \label{sec:distr_BR}

The proposed distributed precoding design is enabled by iterative bi-directional training \cite{Tol19}. A key difference with our previous works \cite{Atz21,Gou20,Atz20} on the unicasting scenario is the addition of a group-specific uplink training resource that entirely eliminates the need for backhaul signaling for CSI exchange in the multi-group multicasting scenario.

At each bi-directional training iteration, the UE-specific effective uplink channel estimation and the effective downlink channel estimation described in Section~\ref{sec:SM_est} are performed, and each UE~$k$ computes its combiner as in \eqref{eq:rxmmse}. The computation of the precoders at each BS requires the cross terms from all the other BSs (see \eqref{eq:dis_bsbf}). To avoid exchanging such cross terms via backhaul signaling, we use an over-the-air signaling scheme similar to that proposed in \cite{Atz21}. Accordingly, each~UE~$k$ transmits $\Y_{k}^{\dl}$ in \eqref{eq:Y_k_dl} after precoding it with $\mu_{k} \v_{k} \v_{k}^{\herm}$, i.e.,
\begin{align}\label{eq:Disultx3}
\X_{k}^{\ulC} \triangleq \sqrt{\beta^{\ulC}}\mu_{k} \v_{k} \v_{k}^{\herm} \Y_{k}^{\dl},
\end{align}
where the scaling factor $\sqrt{\beta^{\ulC}}$ ensures that $\X_{k}^{\ulC}$ complies with the UE transmit power constraint. Then, the received signal at BS~$b$ is given by
\begin{align}
\Y_{b}^{\ulC} & \triangleq \sum_{k \in \setK} \mu_{k} \H_{b,k} \v_{k} \v_{k}^{\herm} \Y_{k}^{\dl} + \Z_{b}^{\ulC}\\
\label{eq:Disulrx3} & = \sqrt{\beta^{\ulC}} \! \sum_{k \in \setK} \! \mu_{k} \H_{b,k} \v_{k} \v_{k}^{\herm} \bigg( \! \sum_{g \in \setG} \! \H_{k}^{\herm} \w_{g}\p_{g}^{\herm} \! + \! \Z_k \! \bigg) \! + \! \Z_{b}^{\ulC},
\end{align}
where $\Z_{b}^{\ulC} \in \Compl^{M \times \tau}$ is the AWGN with i.i.d. $\setC \setN (0, \sigma_{\bs}^{2})$ elements. Consequently, the minimum number of uplink pilot symbols required to obtain $\Y_{b}^{\ulA}$ in \eqref{eq:Y_b_ul1} and $\Y_{b}^{\ulC}$ in \eqref{eq:Disulrx3} with orthogonal pilots at each bi-directional training iteration is $K+G$. Finally, $\Y_{b}^{\ulA}$ and $\Y_{b}^{\ulC}$ are used to compute $\w_{b,g}^{\dis}$ in \eqref{eq:bsbflocal} at the top of the page, with $\D_{\mu} \triangleq \Diag (\mu_1,\ldots,\mu_K)$. Such a precoder replaces $\w_{b,g}^{\star}$ in \eqref{eq:w_bk_i} to achieve global convergence. Note that $\w_{b,g}^{\dis} \to \w_{b,g}^{\star}$ as $\tau \to \infty$. We point out that the additional uplink training resource $\X_{k}^{\ulC}$ conveys the group-specific effective uplink channels of the other BSs instead of the UE-specific effective uplink channels as in the unicasting scenario \cite{Atz21}. The implementation of the best-response distributed precoding design is summarized~in~Algorithm~\ref{alg:disNW}.

\begin{figure}[t!]
\begin{algorithm}[H]
\textbf{Data:} Pilots $\{\p_{k}\}_{k \in \setK}$ (used in UL-1 and UL-3) and $\{\p_{g}\}_{g \in \setG}$ (used in DL). 

\textbf{Initialization:} Combiners $\{\v_{k}\}_{k \in \setK}$.

\textbf{Until} a predefined termination criterion is satisfied, \textbf{do:}
\begin{itemize}
\item[1)] \textbf{UL-1:} Each UE~$k$ transmits~$\X_{k}^{\ulA}$ in $\eqref{eq:X_k_ul1}$; each BS~$b$ receives $\Y_{b}^{\ulA}$ in \eqref{eq:Y_b_ul1}.
\item[2)] \textbf{UL-3:} Each UE~$k$ transmits~$\X_{k}^{\ulC}$ in $\eqref{eq:Disultx3}$; each BS~$b$ receives~$\Y_{b}^{\ulC}$ in \eqref{eq:Disulrx3}.
\item[3)] Each BS~$b$ computes the precoders $\{\w_{b,g}\}_{g \in \setG}$ in \eqref{eq:bsbflocal} and updates them as in \eqref{eq:w_bk_i}. 
\item[4)] \textbf{DL:} Each BS~$b$ transmits a superposition of the pilots $\{\p_{g}\}_{g \in \setG}$ after precoding them with the corresponding precoders $\{\w_{b,g}\}_{g \in \setG}$ as in \eqref{eq:X_b_dl}.
\item[5)] Each UE~$k$ computes its combiner $\v_{k}$ as in \eqref{eq:rxmmse}.
\end{itemize}
\textbf{End}
\caption{(Best-Response Distributed Precoding Design)} \label{alg:disNW}
\end{algorithm} \vspace{-5mm}
\end{figure}

\subsection{Distributed Precoding Design with Group-Specific Pilots} \label{sec:distr_GS}

The best-response distributed precoding design discussed above relies on UE-specific effective uplink channel estimation, which requires the transmission of $K$ orthogonal pilots at each bi-directional training iteration. In the case of scarce training resources, we propose a distributed precoding design based solely on group-specific pilots. This method is obtained by replacing the UE-specific effective uplink channel estimation with its group-specific counterpart described in Section~\ref{sec:SM_est}. Consequently, the minimum number of uplink pilot symbols required to obtain $\Y_{b}^{\ulB}$ in \eqref{eq:Y_b_ul2} and $\Y_{b}^{\ulC}$ in \eqref{eq:Disulrx3} with orthogonal pilots in each bi-directional training iteration is $2G<K+G$. Finally, assuming $\mu_k=1,~\forall k$, $\Y_{b}^{\ulB}$ and $\Y_{b}^{\ulC}$ are used to compute $\w_{b,g}^{\disgs}$ in \eqref{eq:bsbflocalgs} at the top of the page. Such a precoder replaces $\w_{b,g}^{\star}$ in \eqref{eq:w_bk_i} to achieve global convergence.

\section{Numerical Results}

We consider $B=25$~BSs, each equipped with $M=8$~antennas, placed on a square grid with distance between neighboring BSs of $100$~m. Furthermore, $K=32$~UEs, each equipped with $N = 2$~antennas, are divided into $G=8$ multicast groups of $4$ UEs each. Assuming uncorrelated Rayleigh fading, each channel is generated according to $\mathrm{vec}(\H_{b,k}) \sim \setC \setN (\0, \delta_{b,k} \I_{MN})$, where $\delta_{b,k} \triangleq -48-30\log_{10} (d_{b,k})$ [dB] is the large-scale fading coefficient and $d_{b,k}$ [m] is the distance between BS~$b$ and UE~$k$. The maximum transmit power for both the data and the pilots is $\rho_{\bs} = 30$~dBm at the BSs and $\rho_{\ue} = 20$~dBm at the UEs. The AWGN power at the BSs and at the UEs is fixed to $\sigma_{\bs}^{2} = \sigma_{\ue}^{2} = -95$~dBm. Assuming orthogonal pilots for the channel estimation, the antenna-specific, UE-specific, and group-specific uplink channel estimations require $KN$, $K$, and $G$ orthogonal pilots, respectively, whereas the downlink channel estimation requires $G$ orthogonal pilots. In the following we refer to the best-response distributed precoding design (see Section~\ref{sec:distr_BR}) as \textit{distributed best-response}, to the distributed precoding design with group-specific pilots (see Section~\ref{sec:distr_GS}) as \textit{distributed group-specific}, and to the centralized precoding design (see Section~\ref{sec:problem_sumMSE}) as \textit{centralized}.

\begin{figure}[ht]
\centering
\includegraphics[scale=0.42]{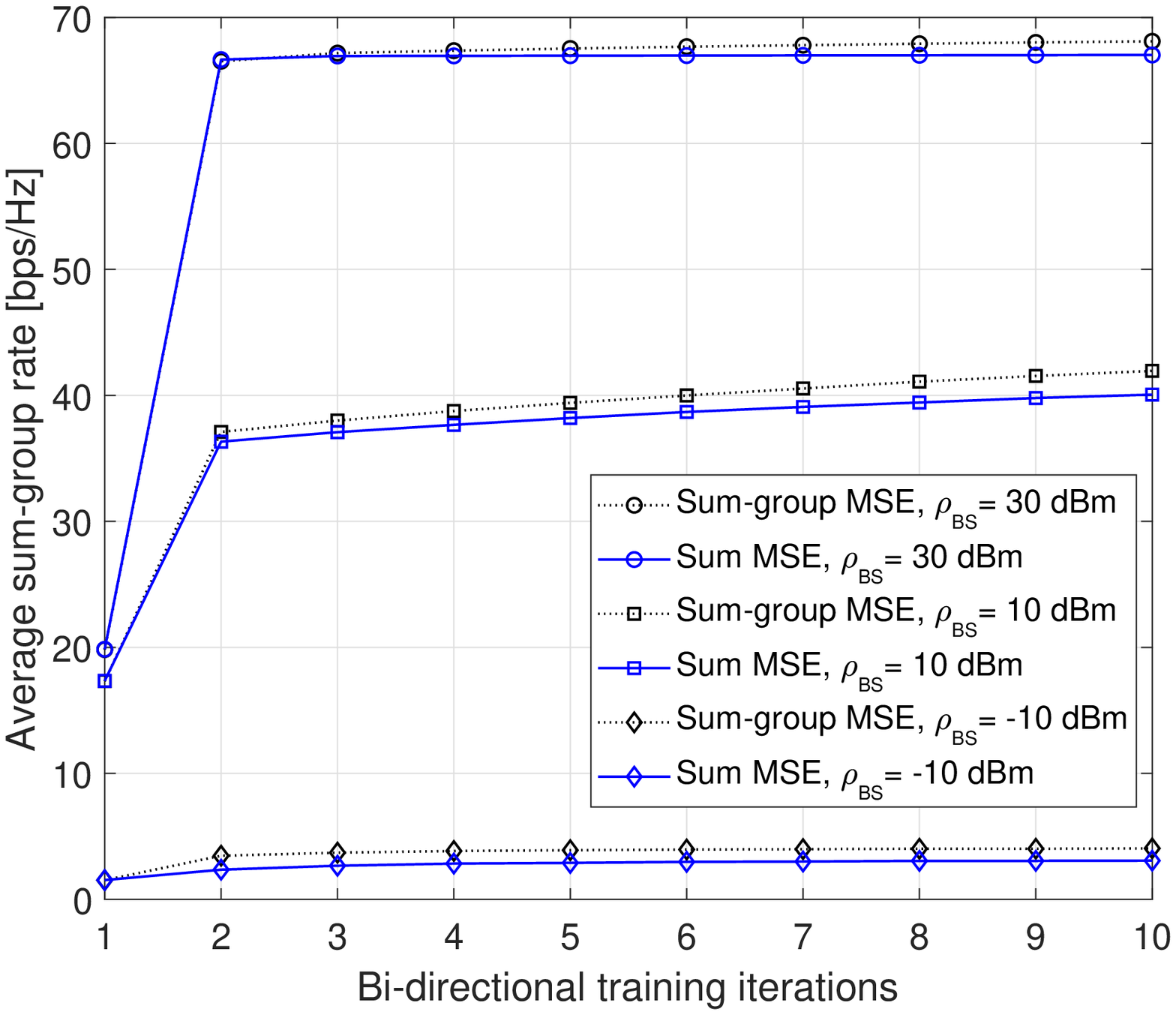}
\vspace{-1mm}
\caption{Average sum-group rate resulting from the sum-group MSE minimization and the sum MSE minimization versus the number of bi-directional training iterations for different values of $\rho_{\bs}$.}
\label{fig:gmseVsSmse}
\vspace{-2mm}
\end{figure}

In Figure~\ref{fig:gmseVsSmse}, we compare the average sum-group rate resulting from the sum-group MSE minimization (see Section~\ref{sec:problem_sum-groupMSE}) and the sum MSE minimization (see Section~\ref{sec:problem_sumMSE}) for different values of $\rho_{\bs}$, with the receive SINRs varying accordingly. As expected, the sum MSE minimization well approximates the sum-group MSE minimization at high SINR.

\begin{figure}[ht]
\centering
\includegraphics[scale=0.42]{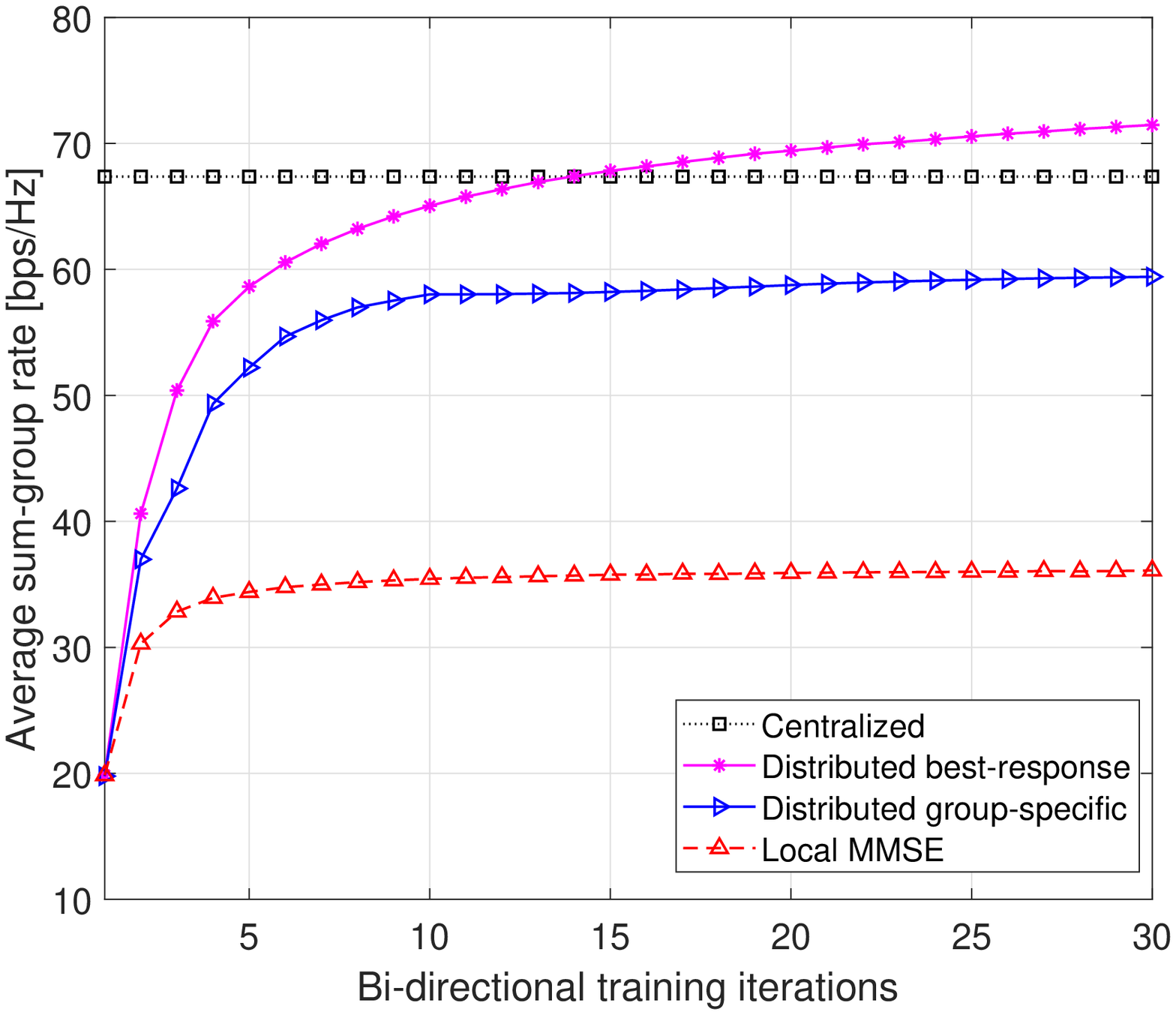}
\vspace{-1mm}
\caption{Average sum-group rate versus the number of bi-directional training iterations.}
\label{fig:rateVsItr}
\vspace{-2mm}
\end{figure}

In Figure~\ref{fig:rateVsItr}, we illustrate the average sum-group rate obtained with the different methods versus the number of bi-directional training iterations. For comparison, we also include the local minimum mean squared error precoding (referred to as \textit{local MMSE}), which is obtained by neglecting the term with $\Y_b^{\ulC}$ in \eqref{eq:bsbflocal}. Remarkably, the \textit{distributed best-response} outperforms all the other schemes, including the \textit{centralized}. As expected, the \textit{distributed group-specific} is worse than the \textit{distributed best-response} as it relies on less accurate local CSI; nonetheless, it still outperforms the \textit{local MMSE} by a large~margin.

\begin{figure}[ht]
\centering
\vspace{-3mm}
\includegraphics[scale=0.42]{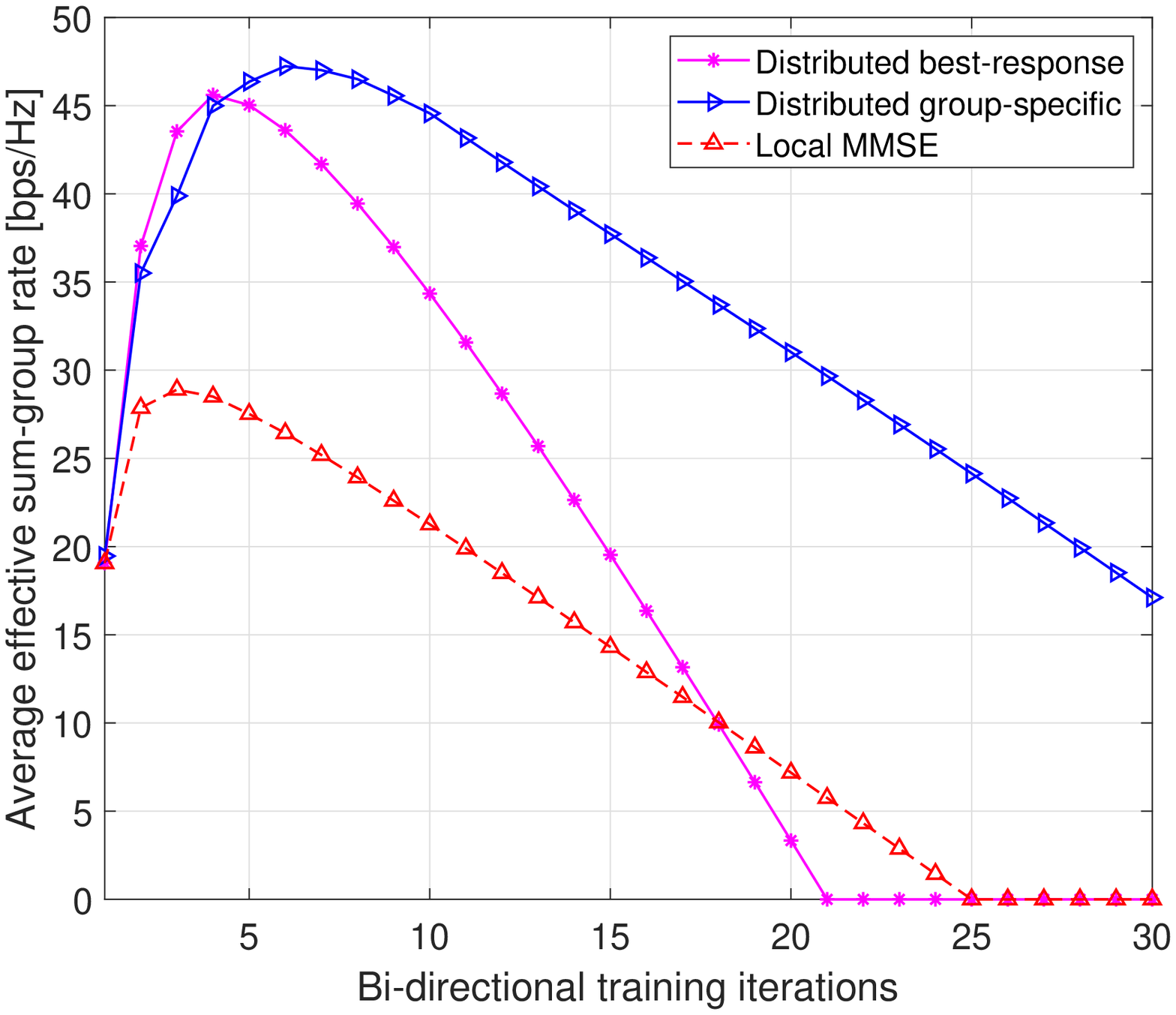}
\vspace{-1mm}
\caption{Average effective sum-group rate versus the number of bi-directional training iterations.}
\label{fig:effrateVsItr}
\vspace{-2mm}
\end{figure}

We now consider the effective sum-group rate calculated as $R_{\mathrm{eff}}^{(i)} \triangleq \big(1 - i \frac{r_{\mathrm{ce}}}{r_\mathrm{tot}}\big)R^{(i)}$, where $i$ is the iteration index, $r_{\mathrm{ce}}$ is the number of pilot symbols used in each bi-directional training iteration, and $r_{\mathrm{tot}}$ is the size of the resource block for the transmission of both the data and the pilots. We neglect the switching time between downlink and uplink signaling in the distributed methods and the backhaul delay for exchanging the cross terms in the \textit{centralized}. In Figure~\ref{fig:effrateVsItr}, we plot the average effective sum-group rate versus the number of bi-directional training iterations for a coherence time of $25$~ms and coherence bandwidth of $40$~kHz, which corresponds to $r_{\mathrm{tot}} = 1000$. The effective sum-group rate of the \textit{distributed best-response} (resp. \textit{distributed group-specific}) with $ r_{\mathrm{ce}} = K+2G$ (resp. $r_{\mathrm{ce}} = 3G$) is around $1.58$ times (resp. $1.63$ times) the sum-group rate obtained with the \textit{local MMSE} with $r_{\mathrm{ce}} = K+G$. Note that the \textit{distributed group-specific} can be implemented even when there are fewer orthogonal pilots than UEs.

\section{Conclusions}

We proposed multi-group multicast precoding designs for cell-free massive MIMO systems. Initially, the sum-group MSE minimization was considered to achieve absolute fairness among the UEs within each multicast group. Subsequently, to simplify the computation and signaling, the sum-group MSE was approximated with the sum MSE. This approximation is accurate in the medium-to-high SINR regime. We utilized iterative bi-directional training with an extra uplink training resource to cooperatively design the precoders at each BS and the combiners at each UE in a distributed fashion. We also proposed a simpler distributed precoding design considering only group-specific pilots, which is useful when the training resources are limited. The proposed distributed precoding designs greatly outperform conventional cell-free massive MIMO precoding techniques such as local MMSE precoding.

\bibliographystyle{IEEEtran}
\bibliography{IEEEabbr,refs}
\end{document}